\documentclass[twocolumn,prl,showpacs, amsmath, amssymb, superscriptaddress,aps,groupedaddress]{revtex4-1}
\usepackage{graphicx}
\usepackage{epstopdf}

\begin{document}
\title{Transmission eigenvalues and the bare conductance in the crossover to Anderson localization}
\author{Zhou Shi}
\affiliation{Department of Physics, Queens College of The City University of New York, Flushing, NY 11367, USA}
\author{Azriel Z. Genack}
\affiliation{Department of Physics, Queens College of The City University of New York, Flushing, NY 11367, USA}
\date{\today}

\begin{abstract}
We measure the field transmission matrix {\it t} for microwave radiation propagating through random waveguides in the crossover to Anderson localization. From these measurements, we determine the dimensionless conductance {\textsl g} and the individual eigenvalues $\tau_n$ of the transmission matrix $tt^\dagger$ whose sum equals {\textsl g}. In diffusive samples, the highest eigenvalue, $\tau_1$, is close to unity corresponding to a transmission of nearly 100$\%$, while for localized waves, the average of $\tau_1$, is nearly equal to {\textsl g}. We find that the spacing between average values of $\ln\tau_n$ is constant and demonstrate that when surface interactions are taken into account it is equal to the inverse of the bare conductance. 
\end{abstract}
\pacs{42.25.Dd, 42.25.Bs, 05.40.-a, 73.23.-b}
\maketitle

More than 50 years ago, Anderson \cite{1} showed that beyond a certain threshold in disorder the electron wavefunction within a material becomes exponentially peaked and diffusion ceases. In the intervening years, localization and its precursors in diffusive samples of enhanced fluctuations and suppressed transmission have been shown to affect every aspect of transport \cite{2,3,4}. Relating transmission in disordered systems to random matrices \cite{5,6,7,8,9,10,11,12} has provided a powerful approach to calculating the scaling and fluctuations of conductance. In this approach, the flow of electrical current through a disordered conductor is assumed to proceed via a set of orthogonal channels on the input and output sides of the sample which are coupled via the field transmission matrix {\it t}. Each channel is a superposition of the {\it N} orthogonal transverse momentum channels supported by the sample leads. Alternatively, {\it t} gives the coupling of the field between different points on opposite surfaces of the sample. The focus of random matrix theory has been the calculation of the dimensionless conductance, {\textsl g}, which is the conductance in units of ($e^2/h$), and equals the sum of the eigenvalues $\tau_n$ of the transmission matrix $tt^\dagger$, $\textsl{g}=\sum_{n=1}^N\tau_n$. However, direct comparisons with measurements of the individual transmission eigenvalues have not been made.

Dorokhov \cite{6} showed that even in conducting samples, the current in most channels would be exponentially small so that the conductance is dominated by ``active'' or ``open'' channels with $\tau_n\ge1/e$. The number of such channels is close to {\textsl g}, $N_{eff}\sim\textsl{g}$ \cite{6,7}. Transmission eigenvalues may be expressed in terms of associated localization lengths, $\xi_n$, $\tau_n=\exp(-L/\xi_n)$ with active channels corresponding to the condition $L\le\xi_n$. Dorokhov found that the average spacing between inverse localization lengths of adjacent eigenchannels in a sample made up of {\it N} parallel chains with weak transverse coupling to neighboring chains was constant and equal to the inverse of the localization length, $1/\xi$. Here, $\xi=N\ell$ is the localization length and $\ell$ is the localization length in a single chain and corresponds to the mean free path [5,6].

Subsequently, Stone {\it et al.} \cite{11} applied a maximum entropy hypothesis to random transfer matrices and found an expression for $\tau_n$ in terms of eigenparameter $\lambda_n$ of the transfer matrix or equivalent parameters $\nu_n$, $\tau_n=1/(1+\lambda_n)=1/\cosh^2(\nu_n/2)$. A “Coulomb gas” model, originally introduced by Dyson \cite{13} to visualize the logarithmic repulsion between the eigenvalues of random Hamiltonian, was extended to describe the repulsion between ``charges'' at positions $\nu_n$. The approximately uniform density of “charges” at $\nu_n$ \cite{10,11} suggested a bimodal distribution for the density of transmission eigenvalues of $\tau_n$   for diffusive waves with peaks near $\tau=0$ and 1 \cite{12,14}. This is confirmed in microscopic calculations made by Nazarov for non-absorbing diffusive samples with perfect leads \cite{15}. 

Recent optical measurements by Vellekoop and Mosk have shown that it is possible to enhance transmission by manipulating an incident optical wavefront to preferentially excite high-transmission channels \cite{16}. Total transmission was enhanced by 44\% and focused intensity by three orders of magnitude by adjusting the phase of the incident optical wavefront reflected from a spatial light modulator (SLM) in response to feedback of transmitted intensity. Popoff and co-workers \cite{17} used an interference technique to measure a portion of the optical field transmission matrix {\it t} using an SLM and CCD camera. They found that the probability density of the singular values of {\it t} corresponded to a quarter-circle law \cite{18,19}. The quarter-circle law arises when the elements of the {\it t} are distributed as independent Gaussian random variables and occurs whenever the dimension of measured {\it t} is smaller than {\textsl g} \cite{20}. Thus these measurements do not reflect correlation in the medium and can not differentiate transmission in samples with different values of {\textsl g}.

In this Letter, we present measurements of the transmission matrix in the crossover to Anderson localization. The transmission matrix is measured for collections of randomly positioned dielectric spheres in a copper tube for which values of {\textsl g} range from 6.9 to 0.17. The dimensionality of the measured matrix matched the number of propagation modes of the empty waveguide. For the most diffusive sample studied, the average value of  $\tau_1$ is 0.93, corresponding to an enhancement of transmission by 900\% over the ensemble average value. The number of “open” eigenchannels with $\tau_n/\tau_1\ge1/e$ with localization lengths greater than the effective sample length is equal to {\textsl g} for diffusive waves. For the most strongly localized waves studied, the average energy transmitted in the highest channel is more than 96$\%$ of the conductance. We find constant spacing between the averages of $\ln\tau_n$ in adjacent channels, $\langle\ln\tau_n\rangle - \langle\ln\tau_{n+1}\rangle\equiv 1/\textsl{g}^{\prime\prime}$, and identical distributions of $\ln\tau_n$ for $n>1$ in both diffusive and localized regimes. We demonstrate that once interactions at the interface are taken into account in terms of effective sample length and localization length, $L_{eff}$ and $\xi$, respectively, the constant spacing of values of $\langle\ln\tau_n\rangle$ is equal to the inverse of the bare conductance, giving $\textsl{g}^{\prime\prime}=\textsl{g}_0=\xi/L_{eff}$. 

Because the phenomena we seek to explore depend only upon {\textsl g}, the particular sample studied is immaterial. Measurements were carried out on well characterized samples \cite{21} for which we could access values of {\textsl g} through the localization transition. The experiment setup is sketched in Fig. 1a. The samples are composed of alumina spheres with a diameter of 0.95 cm and refractive index 3.14 embedded in Styrofoam shells and contained within a copper tube of diameter 7.3 cm. The field transmission matrix was measured in samples with lengths $L=23$, 40 and 61 cm over the frequency ranges 10-10.24 GHz and 14.70-14.94 GHz with use of a vector network analyzer. The wave is localized in the lower frequency range and diffusive in higher frequencies. The numbers of transverse waveguide modes supported in the lower and upper frequency ranges are $N\sim30$ and $N\sim66$, respectively. Measurements are made for two linearly polarized components of radiation along the horizontal and vertical orientations of the source and detector antennas between $N/2$ points on a grid on the input and output surfaces of the sample tube in each frequency range. New statistically equivalent samples are produced by momentarily rotating and vibrating the copper tube after the full matrix {\it t} is recorded in both frequency ranges over a period of two days. The impact of absorption upon the statistics of {\it t} is removed by compensating the field transformed into the time domain for the exponential decay due to absorption and then transforming back into the frequency domain. For the shortest sample studied for which $\textsl{g}=6.9$, the transmittance is reduced by 25\% due to absorption. Measurements were made in a total of 23 sample configurations for $L=23$ cm and 6 configurations for $L=40$ and 61 cm.  

Speckle patterns for two different positions {\it a} and $a^\prime$ of the source antenna in a random realization at a single frequency are shown for a diffusive sample with $L=23$ cm, $\textsl{g}=6.9$ and for a localized sample with $L=40$ cm, $\textsl{g}=0.37$ in Fig. 1b.
\begin{figure}[htc]
\includegraphics[width=3.2in]{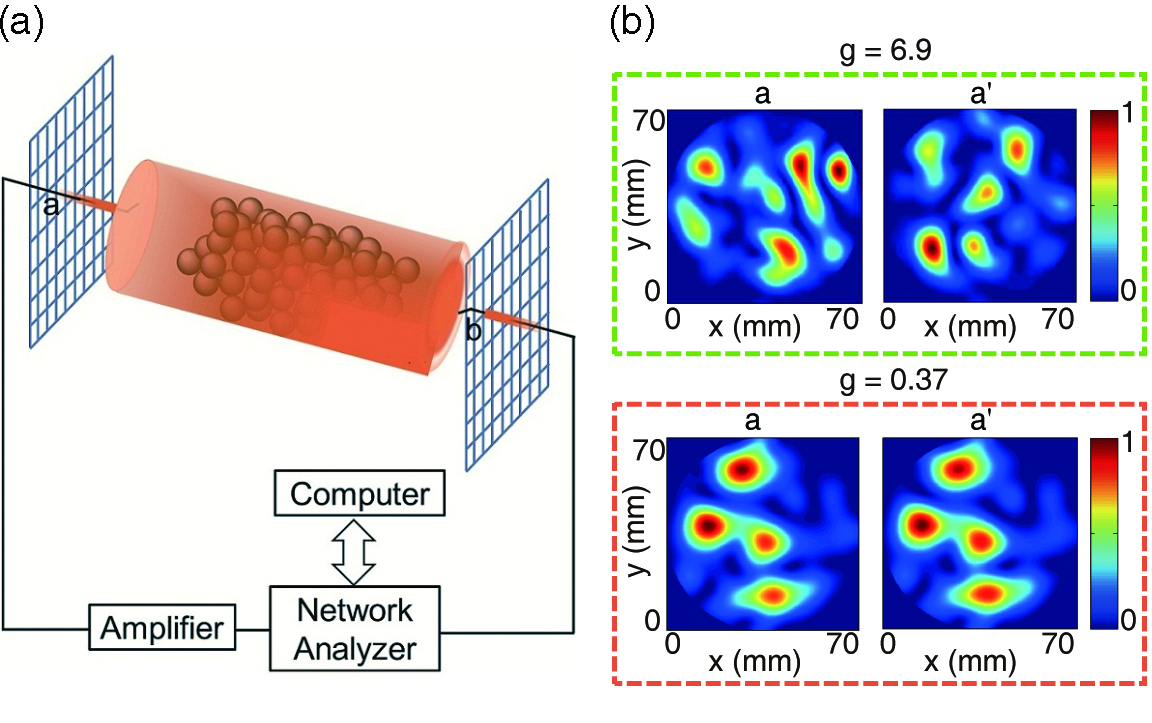}
\caption{\label{Fig1}(Color online) (a) Schematic experiment setup. Source and detector wire antennas are mounted on a two-dimensional translation stage, so they could move freely on a two- dimensional grid. Measurements are made for two orthogonal linear polarizations by rotating the wire antennas. (b) Speckle patterns from two different source positions $\textit{a}$ and $\textit{a}$$^\prime$ at a single frequency in one random realization for diffusive waves with $\textsl{g}=6.9$, (upper row) and localized waves with $\textsl{g}=0.37$, (lower row).}
\end{figure}
Characteristically, the speckle patterns for diffusive waves are quite different, while the patterns for localized waves are nearly the same. The speckle patterns are different for diffusive waves because the transmitted field is a random admixture of many transmission eigenchannels. In contrast, the ratio of successive transmission eigenvalues for localized waves is large and a single channel typically dominates transmission. This is confirmed in the spectra of the first few transmission eigenvalues for diffusive and localized waves shown in Fig. 2. Also shown in Fig. 2 is the transmittance, which is the sum of the flux between all $N^2$ pairs of incident and outgoing channels, $T=\sum_{a,b=1}^N|t_{ab}|^2$. Approximately 7 highly transmitting channels with values of $\tau_n/\tau_1\ge1/e$ are seen to contribute to the transmittance {\it T} with average dimensionless conductance $\textsl{g}=\langle T\rangle=6.9$. In contrast, in the sample with $\textsl{g}=0.37$, $\langle\tau_2\rangle/\langle\tau_1\rangle\sim0.088$, and transmission is dominated by the first eigenchannel for most positions of the source. 
\begin{figure}[htc]
\includegraphics[width=2.8in]{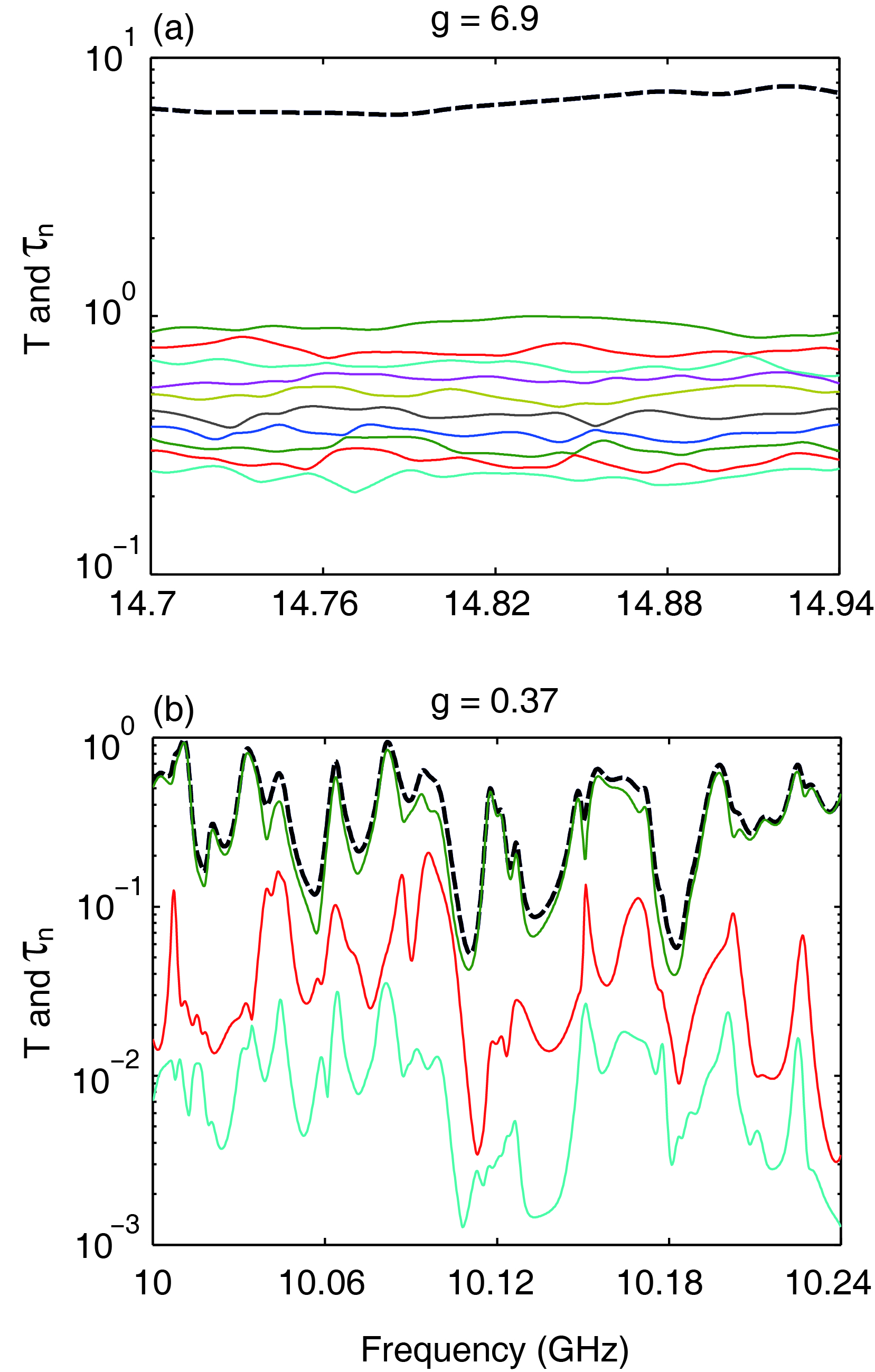}
\caption{\label{Fig2}(Color online) Spectra of the transmittance $\textit{T}$ and transmission eigenvalues $\tau_n$ for (a) diffusive sample of {\it L}=23 cm with $\textsl{g}=6.9$ and (b) localized sample of {\it L}=40 cm with $\textsl{g}=0.37$. The black dashed line gives $T=\sum_{a,b=1}^N|t_{ab}|^2=\sum_{n=1}^N\tau_n$ and the solid lines are spectra of $\tau_n$.}
\end{figure}
The average value of $\tau_1$ of 0.93 for diffusive waves is nearly a factor of 10 greater than the average value of the total transmission for an incident wave from a source at point {\it a}, $\langle T_a\rangle=\langle T\rangle/N=\textsl{g}/N\sim0.1$. For localized waves, $\tau_1\sim T$, so transmission is enhanced over $\langle T_a\rangle$ by a factor nearly equal to $N =30$. 

The relative values of conductance for samples of different length and waves at different frequency are given by the ratio of the ensemble average of the sum of intensity for all pairs of incident and output points and the same measurement in a tube emptied of scatterers. The absolute transmittance is found using a normalizing factor obtained by equating $\langle T\rangle=\textsl{g}$ with the bare conductance $\textsl{g}_0=\textsl{g}^{\prime\prime}=6.9$ in the most diffusive sample. This sample is far from the localization threshold and corrections to {\textsl g} due to renormalization are expected to be negligible. We show below that the scaling $\textsl{g}^{\prime\prime}$ provides an accurate determination of $\textsl{g}_0$. 

The statistics of  $\ln\tau_n$ are presented in Fig. 3 for ensembles with different values of {\textsl g}. The probability density of individual $\ln\tau_n$ of the first few eigenchannels and their contribution to the overall probability density of $P(\ln\tau)=\sum_{n=1}^NP(\ln\tau_n)$ for the most diffusive sample of $\textsl{g}=6.9$ is presented in Fig. 3a. Aside from the fall of the probability density $P(\ln\tau)$ near $\ln\tau\sim0$, which reflects the restriction $\tau_1\le1$, $P(\ln\tau)$ is nearly constant with ripples spaced by $1/\textsl{g}^{\prime\prime}$ \cite{22,23}. We find that the density of $\ln\tau_n$ for $n>1$ is a Gaussian function peaked at $\langle\ln\tau_n\rangle$ with width independent of the index {\it n} \cite{24}. The density of $u=u_n\equiv\ln\tau_n-\langle\ln\tau_n\rangle$ for {\it n} between 2 and 10 is shown in Fig. 3b and compared with a Gaussian distribution. The regular spacing of $\langle\ln\tau_n\rangle$ seen in Fig. 3c and the nearly identical distribution of $\ln\tau_n$ for $n>1$ is reminiscent of a crystalline lattice arrangement of atoms at a finite temperature. 

\begin{figure}[htc]
\includegraphics[width=2.8in]{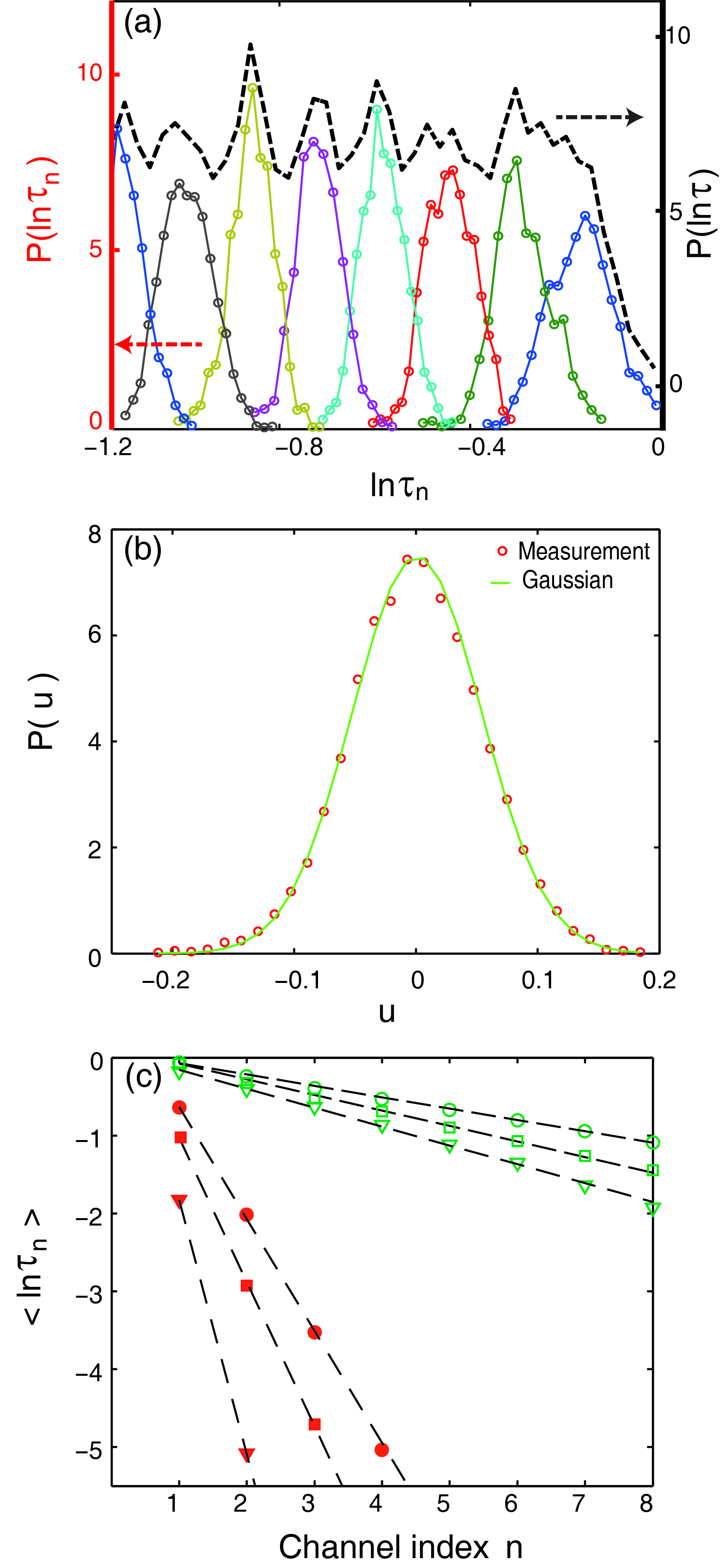}
\caption{\label{Fig3}(Color online) ``Crystallization'' of transmission eigenvalues. (a) Probability density of $\ln\tau_n$ (lower curves) and the density of $\ln\tau$ (top dash curve), $P(\ln\tau)=\sum_{n}P(\ln\tau_n)$ for diffusive sample with $\textsl{g}=6.9$. (b) Probability density of $u=u_n\equiv\ln\tau_n-\langle\ln\tau_n\rangle$ for $n$ between 2 and 10, for the same sample as in (a) compared with a Gaussian distribution. (c) Variation of $\langle\ln\tau_n\rangle$ with channel index {\it n} for sample lengths $L=23$ (circle), 40 (square) and 61 (triangle) cm for both diffusive(green open symbols) and localized(red solid symbols) waves fitted, respectively, with black dash lines.}
\end{figure}

The slope of the curve in Fig. 3c, is the spacing between successive values of $\langle\ln\tau_n\rangle$, $1/\textsl{g}^{\prime\prime}$. For diffusive samples, for which there are several channels with $\tau_n/\tau_1\ge1/e$, the number of such channels is equal to $\textsl{g}^{\prime\prime}$. Dorokhov \cite{6} predicted that $\textsl{g}^{\prime\prime}$ is equal to the bare conductance, $\textsl{g}^{\prime\prime}=\textsl{g}_0=N\ell/L$. But the bare conductance $\textsl{g}_0$ should be influenced by wave interactions at the sample interface, as is the case for transmission for a single incident channel \cite{25,26,27}. The impact of the interface upon transmission can be found by considering the angular dependence of transmission.

Measurements of total transmission in diffusive samples are well described by the expression, $T(\theta) = (z_p\cos\theta+z_b)/(L+2 z_b)=(z_p\cos\theta+z_b)/L_{eff}$, where $\theta$ is the angle between the normal and the wave as it penetrates into the sample \cite{27}. This expression is obtained from a model in which the incident wave is replaced by an isotropic source at a distance of travel from the interface,  $z_p$, at a depth, $z_p\cos\theta$ as illustrated in Fig. 4.
\begin{figure}[!tp]
\includegraphics[width=2.65in]{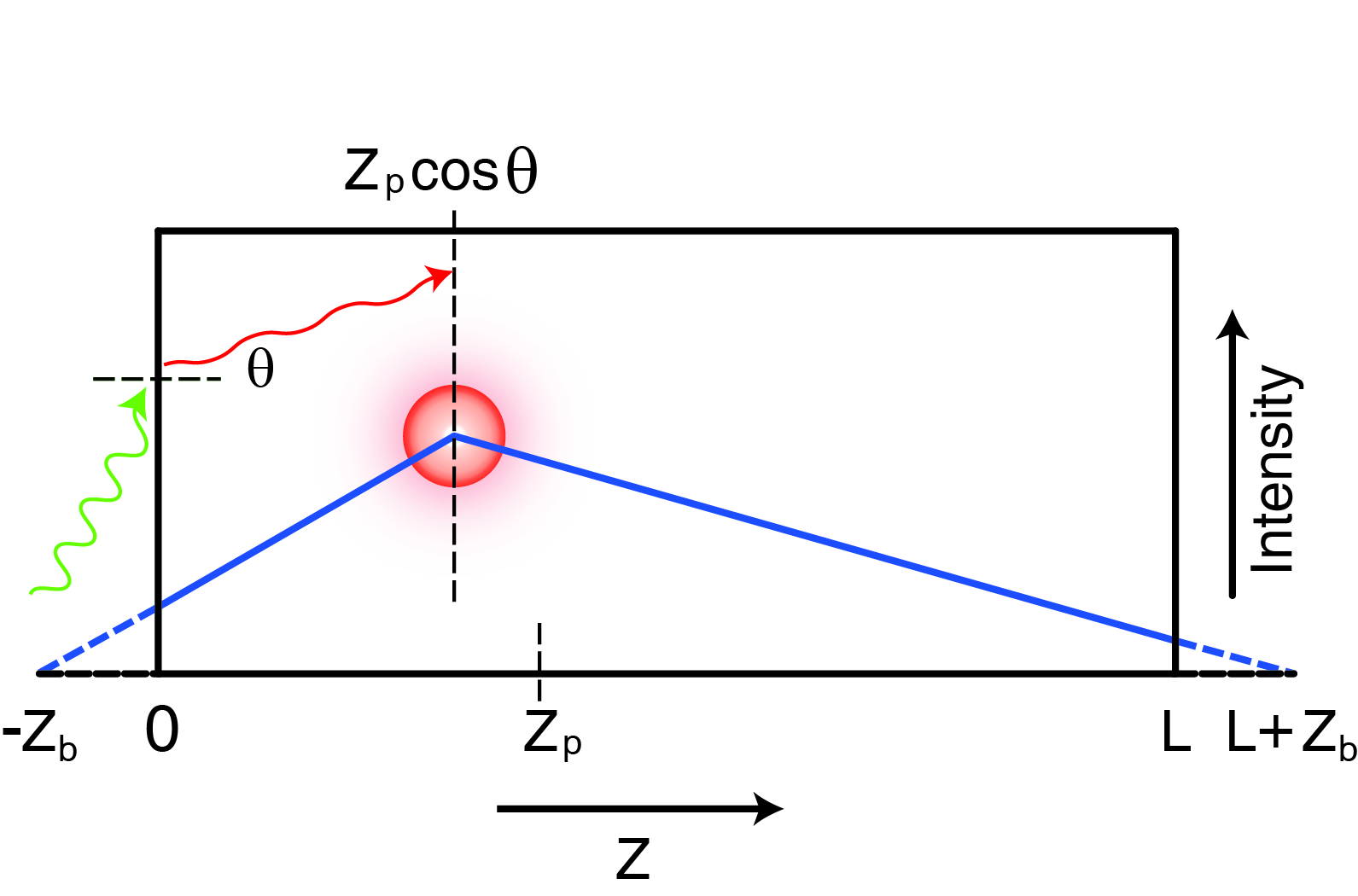}
\caption{\label{Fig4}(Color online) Illustration of the photon diffusion model for the spatial variation of intensity, $I(z)$.}
\end{figure} 

Diffusion within the sample gives a constant gradient of intensity to the right and left of the effective randomization depth $z_p\cos\theta$, which extrapolates to zero at a length $z_b$ beyond the sample boundary. The gradients of intensity at the input and output surfaces give the reflection and transmission coefficients, which together with the condition, $T + R = 1$, gives the expression for $T(\theta)$ above. The linear falloff of intensity near the surface boundaries should hold even for localized waves since the diffusion coefficient varies with depth into the sample but is hardly renormalized near the sample boundaries \cite{28}. Since $z_p$ and $z_b$ are proportional to the mean free path, $\ell$, averaging over all incident angles gives, $\textsl{g}_0=\eta N\ell/(L+2z_b)=\eta N\ell/L_{eff}=\xi/L_{eff}$. Here $\eta$ is independent of {\it L} and includes the effects of reduced flux into the sample due to external reflection of the incident wave at the interface, and enhanced internal reflection. These effects tend to cancel in transmission so that $\eta\sim1$. The values of $z_b$ of 13 and 6 cm for the diffusive and localized samples, respectively, are obtained by fitting the diffusion model to the measured time of flight distribution \cite{29,30}. Hypothesizing that $\textsl{g}^{\prime\prime}=\textsl{g}_0$, the product $\textsl{g}^{\prime\prime}L_{eff}$ should give a constant length, which we can identify as $\xi$. We find this product to be essentially constant for measurements carried out in the three sample lengths in each of the two frequency ranges as shown in Fig. 5.

\begin{figure}[htc] 
\includegraphics[width=2.5in]{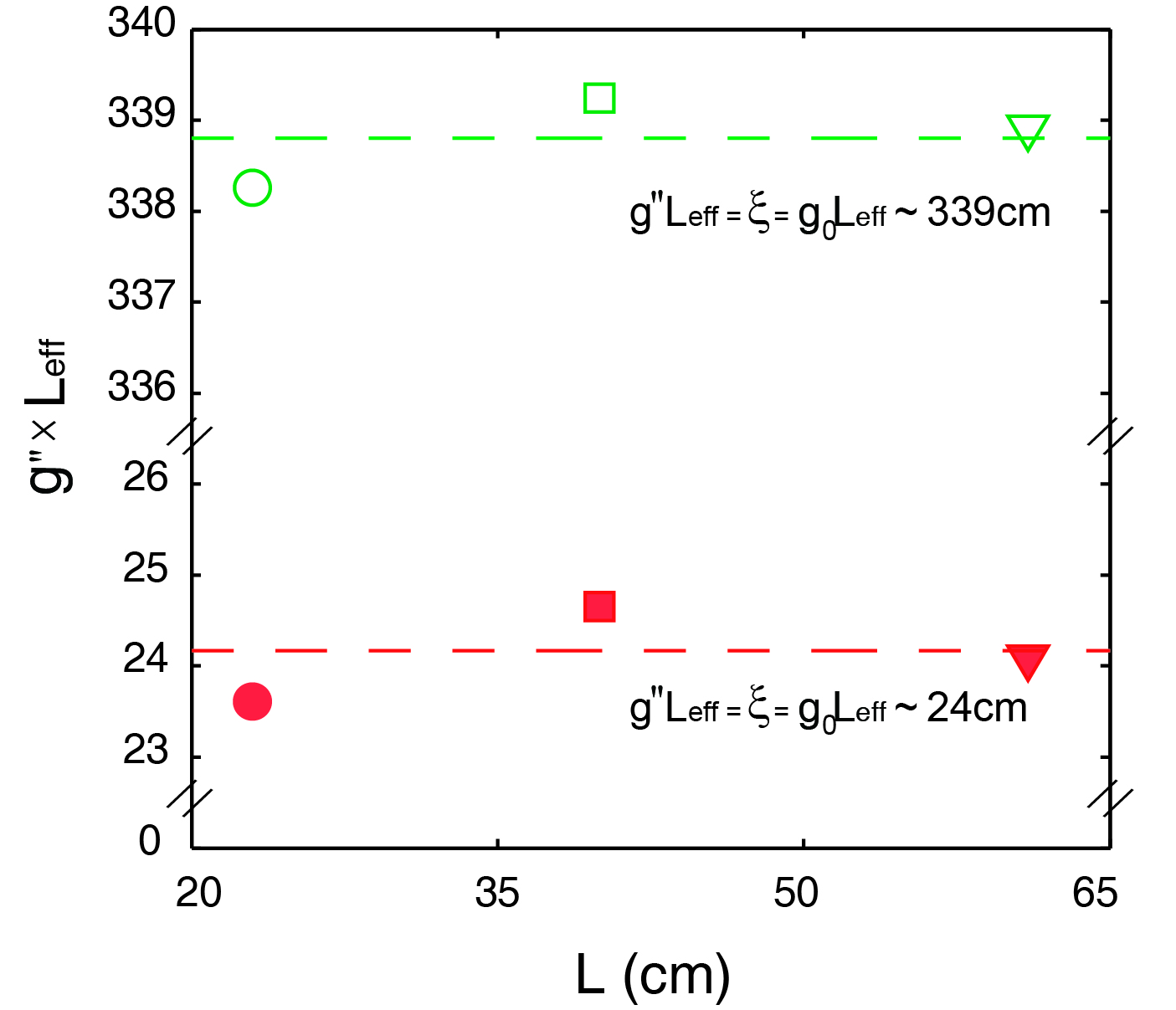}
\caption{\label{Fig5} Identification of $\textsl{g}^{\prime\prime}$ with the bare conductance $\textsl{g}_0$. The constant products of $\textsl{g}^{\prime\prime}L_{eff}$ for three different lengths for both diffusive and localized samples give the localization length $\xi$ in the two frequency ranges.}
\end{figure}

The nearly uniform density of $P(\ln\tau)$, seen in Fig. 3a, which is associated with the equal spacing between successive values of $\langle\ln\tau_n\rangle$ is consistent with calculations of Dorokhov \cite{5,6}. These results correspond to a nearly uniform probability density $P(\ln\tau)=\textsl{g}$ for diffusive waves, apart from a small modulation due to the peaking of this density at $\ln\tau=\langle\ln\tau_n\rangle$. This leads to the probability density, $P(\tau)=P(\ln\tau)\frac{d\ln\tau}{d\tau}=\textsl{g}/\tau$. This distribution has a single peak at low values of $\tau$ in contrast to later prediction of a bimodal distribution for $\tau$ for diffusive waves which has a second peak near unity. This difference may be a consequence of differences in the circumstances of our measurements as compared to the assumptions made in many analytical calculations. In our measurements, the wave is a vector, the sample is not in the extreme diffusive limit of $\textsl{g}\gg1$, {\it t} couples measurements on a grid of points, and surface interactions are included. Another difference between measurements and calculations is that flux is not conserved in our sample so that the measured transmission matrix is not Hermitian. However, the spacing of transmission eigenvalues does not appear to be affected appreciably by moderate absorption. In the most diffusive sample studied with $\textsl{g}=6.9$, the spacing between adjacent values of $\langle\ln\tau_n\rangle$ is uniform and remains so when the field spectra are compensated for absorption. Although transmission is suppressed by 25\% by absorption, the average spacing is only reduced by $\sim2\%$. The insensitivity of the statistics of transmission eigenvalues to absorption parallels the insensitivity to absorption of the variance of total transmission normalized by its ensemble average, which is suppressed by $\sim1.6\%$ \cite{21}. 

In conclusion, we found the transmission eigenvalues $\tau_n$ and the optical transmittance {\it T} for microwave radiation propagating through random waveguides for diffusive and localized waves. We show that the scaling of $\langle\tau_n\rangle$ and $\textsl{g}$ is determined by the bare conductance, $\textsl{g}_0=\xi/L_{eff}$, which is strongly affected by surface interactions. These measurements make it possible to directly test random matrix theories based on different assumptions and provide a systematic basis for controlling the transmitted field for applications in communications, imaging and lasing in complex systems.   

We thank Jing Wang, Matthieu Davy and Chushun Tian for stimulating discussions and Howard Rose for advice on the operation of the experiment. The research was supported by the NSF under Grant Nos. DMR-0907285 and DMR-0958772 MRI-R2.

\end{document}